\documentclass[aps,prb,reprint,longbibliography,superscriptaddress]{revtex4-2}
\usepackage{mathrsfs}
\usepackage{amsmath,gensymb}
\usepackage{amsfonts}
\usepackage{amssymb}
\usepackage{amsthm}
\usepackage{graphicx}
\usepackage{natbib}
\usepackage{color}
\usepackage{hyperref}
\usepackage{bm}
\usepackage[caption=false]{subfig}
\usepackage{verbatim}
\usepackage[normalem]{ulem}

\begin{document}
\title{Theory of planar quasiballistic Josephson junctions}

\author{G.A. Bobkov}
\affiliation{Moscow Institute of Physics and Technology, Dolgoprudny, 141700 Moscow region, Russia}

\author{I.V. Bobkova}
\affiliation{Moscow Institute of Physics and Technology, Dolgoprudny, 141700 Moscow region, Russia}
\affiliation{HSE University, 101000 Moscow, Russia}

\author{A.M. Bobkov}
\affiliation{Moscow Institute of Physics and Technology, Dolgoprudny, 141700 Moscow region, Russia}

\author{K.B. Polevoy}
\affiliation{Moscow Institute of Physics and Technology, Dolgoprudny, 141700 Moscow region, Russia}
\affiliation{All-Russian Research Institute of Automatics n.a. N.L. Dukhov (VNIIA),
127030, Moscow, Russia}




\author{V.~S.~Stolyarov}
\affiliation{Moscow Institute of Physics and Technology, Dolgoprudny, 141700 Moscow region, Russia}
\affiliation{All-Russian Research Institute of Automatics n.a. N.L. Dukhov (VNIIA),
127030, Moscow, Russia}


\begin{abstract}
We develop a theoretical framework for planar quasiballistic Josephson junctions, where multiple superconducting leads are coupled through a large, nearly ballistic normal metal crystal. Our approach, based on quasiclassical Eilenberger equations, accounts for the dominant role of electron reflections from the crystal surfaces or single impurities, a mechanism distinct from both purely ballistic and diffusive limits. We calculate the critical current between superconducting leads for various geometries, examining its dependence on temperature and magnetic field. Crucially, we demonstrate that in multiterminal setups, the junctions are not independent but form a strongly coupled system. The theory successfully explains key experimental observations from a companion work [Polevoy et al., Phys. Rev. Lett. 137, 166001 (2026)], including a non-monotonic dependence of the critical current on the interlayer length, providing a foundation for designing and understanding complex multi-terminal Josephson systems.
\end{abstract}

\maketitle

\section{Introduction}

There is currently a growing interest in complex Josephson structures, where a non-superconducting region is proximitized by several superconducting electrodes \cite{Pfeffer2014,Strambini2016,Vischi2017,Cohen2018,Graziano2020,Pankratova2020,Graziano2022,Kozler2023,Draelos2019,Arnault2021,Huang2022}. In such devices, the interplay of supercurrent flow involving different superconducting leads can be highly non-trivial. In particular, the energy spectrum of a multi-terminal
Josephson device can serve as an analog of the band structure of a crystal, where the
superconducting phases play the role of quasi-momenta of an arbitrary dimension \cite{vanHeck2014,Yokoyama2015}. Moreover, it was reported that such artificial band structure can provide a platform for implementation and investigation of the topological matter \cite{Riwar2016,Gavensky2023}.

Several research groups have successfully implemented mesoscopic Josephson junctions (JJs) featuring three or four terminals in various settings, including  diffusive metallic junctions \cite{Pfeffer2014,Strambini2016,Vischi2017}, hybrid semiconductor-superconductor heterostructures \cite{Cohen2018,Graziano2020,Pankratova2020,Graziano2022} and topological materials \cite{Kozler2023}. Ballistic JJs hold particular promise for the realization of multi-terminal Josephson systems, as they can facilitate effective Josephson coupling between superconducting leads that are several microns apart. To date, experimental implementations and investigations of such ballistic multi-terminal systems have been conducted using graphene \cite{Draelos2019,Arnault2021,Huang2022}. 

\begin{figure*}[t]
\includegraphics[width=160mm]{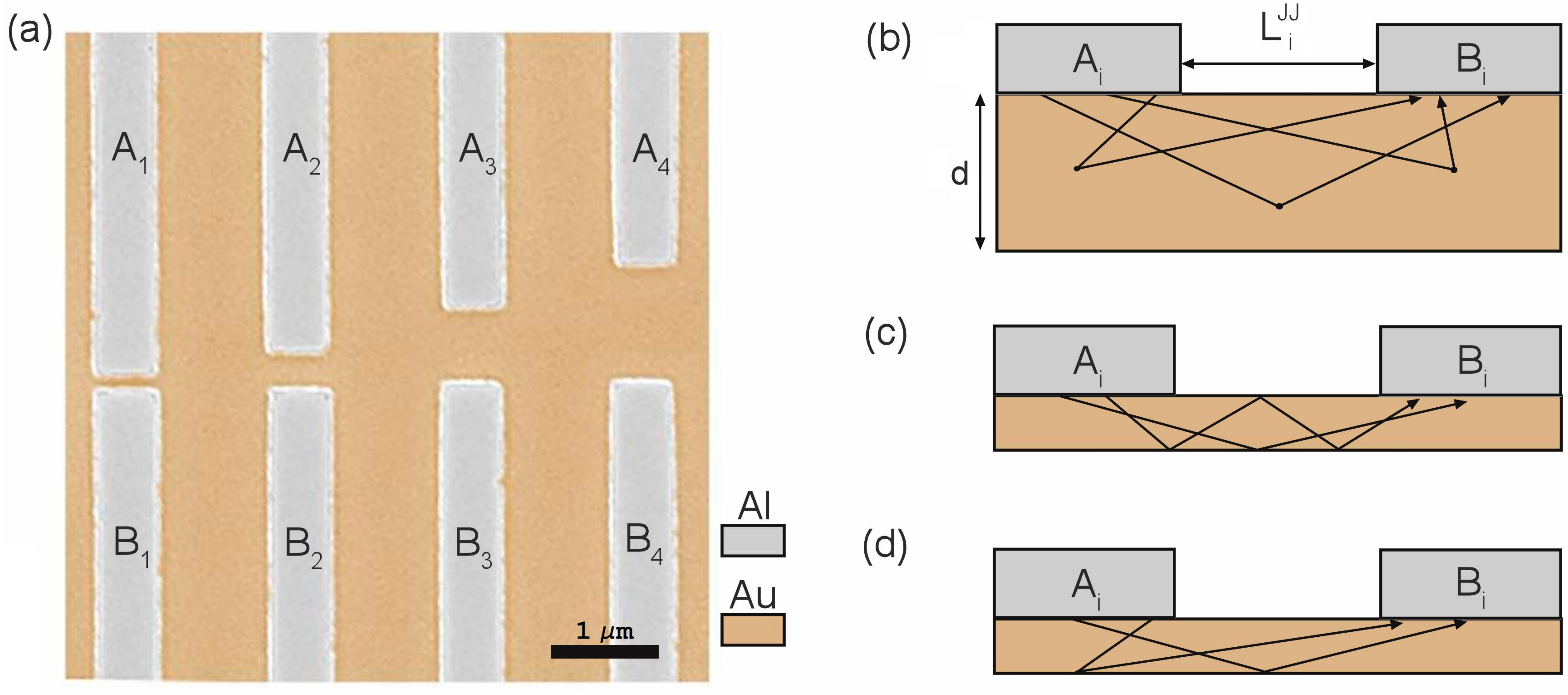}
\caption{(a) Colored SEM image of the studied complex system of Josephson junctions (JJs) fabricated by placing the Al superconducting electrodes on top of the Au monocrystalline sample. (b)-(d) Different models used in the work for description of the quasi-ballistic planar JJs. The main contribution to the electric current is given via (b) single scattering event at an impurity in the bulk of the semi-infinite N crystal; (c) multiple specular reflections at the N layer surfaces; (d) single diffusive reflection at the bottom N layer surface.}
 \label{fig:setup}
\end{figure*}

In a companion work \cite{Polevoy2025_joint}, another experimental implementation of a multi-terminal Josephson system on a quasi-ballistic Au single crystal is presented. The SEM image of the experimental setup is presented in Fig.~\ref{fig:setup}(a). Such
SN/N/SN systems, that is, planar Josephson junctions with superconducting leads (S) deposited on top of the normal metal (N), are of particular interest for realization of multi-terminal Josephson devices, but at the same time, the corresponding theoretical description is still lacking. By now, in almost all works planar SN/N/SN structures have been described theoretically in the diffusive limit \cite{Golubov2004,Soloviev2021,Ruzhickiy2023,Bosboom2021,Marychev2020,2024_Bakurskiy,Polevoy_2026,Vodolazov_2026,Bakurskiy_2026}. In the experimental system under study \cite{Polevoy2025_joint} the low-temperature mean free path $l=680$nm, the ballistic superconducting coherence length $\xi=v_F/2\pi T_c \approx 1.7 \mu$m (where $v_F$ is the electron Fermi velocity and $T_c$ is the superconducting critical temperature of the leads) and the diffusive coherence length $\xi_d = \sqrt{l \xi}$ are of the same order of magnitude. Therefore, the condition of the diffusive limit $l \ll \xi_d$ is not fulfilled. Moreover, the thickness of the Au single crystal $d \approx 80$nm is also much smaller than the mean free path.  Thus, the description within the diffusive limit is not applicable. From the other hand, for the planar geometry under consideration  a description within the framework of purely ballistic theories \cite{Kulik1969,Ishii1970,Bardeen1972,Kupriyanov1981,Galaktionov2002,Golubov2004} is also inappropriate, since none of the ballistic trajectories leaving one superconducting lead enters the other. Only trajectories on which an electron experiences single or multiple reflections from the bottom surface of the Au crystal or from impurities can contribute to the Josephson current, see Fig.~\ref{fig:setup}(b)-(d). Similar ballistic planar junctions with mirror reflections from the surfaces of the normal interlayer were only considered for translationally invariant infinite superconducting leads \cite{Kuprianov1986,Kuprianov1988}, which does not allow for the description of the multi-terminal Josephson structures. Here we present a theoretical framework to describe the strongly coupled quasi-ballistic planar Josephson structures fabricated on the surface of a large crystal. The comparison of the results of the theory with the experimental data was discussed in the companion work \cite{Polevoy2025_joint}. 

The paper is organized as follows. In Sec.~\ref{models} we describe theoretical models used for the calculation of the Josephson properties of the planar quasi-ballistic JJs on top of crystals of different thicknesses. In Sec.~\ref{greens_functions} the quasiclassical Green's functions approach is formulated. In Sec.~\ref{current} the Josephson current through a single planar quasi-ballistic JJ is discussed and in Sec.~\ref{multiterminal} the results for the multi-terminal systems in the framework of the different studied models are presented. The main results of the work and conclusions are summarized in Sec.~\ref{conclusions}.

\section{Models of a quasiballistic planar JJ on an infinite crystal}
\label{models}


As it was mentioned in the introduction, for the
planar SN/N/SN geometry a description within the framework of purely ballistic theories is inappropriate, since none of the ballistic trajectories leaving one superconducting lead enters the other. The current is carried by trajectories on which an electron experiences single or multiple reflections from the surfaces of the Au crystal or from impurities in the bulk of the crystal. The choice of the most suitable model is determined primarily by the thickness $d$ of the normal crystal, see Fig.~\ref{fig:setup}(b)-(d), but in any case for considered quasiballistic system we assume $L_i \lesssim l$. Here $L_i$ is the length of the interlayer between the leads.

If the thickness of the normal metal is large compared to the mean free path and the coherence length, $d \gtrsim (\xi,l)$ then the main contribution to the Josephson current is given by the impurity scattering in the normal layer. The corresponding scattering processes are sketched in Fig.~\ref{fig:setup}(b). In this case the crystal can be treated as semi-infinite. In considering this model we take into account only single processes of scattering on impurities. Strictly speaking, the applicability of this simplification is determined by the parameter $(\xi/l)e^{-L_i/\xi} \ll 1$, as it is shown in Sec.~\ref{bulk_imp_model}.

In the opposite case of the thin normal crystal with $d \ll (\xi,l)$ the main contribution to the current is provided via the mechanism of reflections from its surfaces, and the scattering at the impurities in its bulk is disregarded. Depending on the roughness of these surfaces, the reflection may contain both a specular and a diffusive component. It is generally not possible to take into account the details of the roughness at the boundaries, so we restrict ourselves by two limiting cases that allow analytical consideration. The first limit is the multiple specular reflections at the  upper and lower surfaces of the normal layer, see Fig.~\ref{fig:setup}(c). The considered approximation is applicable if the roughness of the normal layer surfaces is small and $(d,L_i) < l$. This model is a generalization of the previously considered effectively 1D model of translationally invariant ballistic planar junctions with specular reflections from the surfaces for the case of an arbitrary geometry of the superconducting leads\cite{Kuprianov1986,Kuprianov1988}.

The second easily solvable limit for the thin normal crystal is a single isotropic diffusive scattering from the bottom surface of the normal crystal, see Fig.~\ref{fig:setup}(d). In this model the multiple diffusive reflections from the upper and bottom surfaces are disregarded. This approximation works well if $\xi<d<l$ or if $L_i \lesssim 2d$ (see Sec.~\ref{film_diffusive_model}).

Below we develop the quasiclassical Green's function approach for calculation of the Josephson current of complex Josephson systems with an arbitrary geometry of the superconducting leads in the framework of the three described models.

\section{Quasiclassical Green's functions formalism}
\label{greens_functions}

\subsection{Quasiclassical description of a low-transparency quasi-ballistic interface}

The calculation of the Josephson current in the framework of all the considered models is based on the Eilenberger equation \cite{Eilenberger1968,Larkin1968} for the quasiclassical Green's function $\check g(\bm n, \bm r, \omega)$, which is a matrix $2 \times 2$ in the particle-hole space and depends on the electron trajectory direction $\bm n =\bm v_F/v_F$, spatial coordinate $\bm r$ and the fermionic Matsubara frequency $\omega = \pi T (2n+1)$, where $n$ is an integer number:
\begin{align}
    [i\omega \tau_z+\left(
\begin{array}{cc}
0 & \Delta \\
-\Delta^* & 0
\end{array}
\right)+\frac{i}{\tau}\langle \check g \rangle,~\check g]+i \bm v_F\bm \nabla \check g = 0 ,
\label{eq:eilenberger}
\end{align}
where $\tau_i$ for $i=x,y,z$ are the Pauli matrices in the particle-hole space, $\Delta$ is the superconducting order parameter, which is nonzero only in the leads, $\langle \check g \rangle$ is the quasiclassical Green's function averaged over all trajectories, $\tau$ is the impurity mean free time and the term $-i\langle \check g \rangle/\tau$ represents the impurity self-energy in the self-consistent Born approximation. The Eilenberger equation (\ref{eq:eilenberger}) should be supplemented by the normalization condition:
\begin{align}
    \check g^2=1.
    \label{normalization}
\end{align}
The explicit matrix structure of $\check g$ in the particle-hole space takes the form:
\begin{align}
    \check g=\left(
\begin{array}{cc}
g & f \\
\tilde f & \tilde g
\end{array}
\right),
\end{align}
where $g$, $\tilde g$ are normal Green's functions and $f$, $\tilde f$ are anomalous Green's functions. From the normalization condition it follows that $\tilde g = - g$ and $f\tilde f+g^2=1$. Assuming that transmission $D$ of the interface between superconductor and normal metal (S/N) is small, $D\ll 1$, which is relevant to the experimental situation, the boundary conditions at the S/N interface take the form \cite{Zaitsev1984}:
\begin{align}
    \check g_N^{out}-\check g_N^{in}=\frac{D}{2}[\check g_N^0,\check g_S^0],
    \label{eq:bc}
\end{align} 
where $\check g_N^{in(out)}$ is the Green's function incoming to (outgoing from) the N side of the S/N interface, and $\check g_{N(S)}^0$ is the Green's function in the normal metal (superconductor) at $D=0$. The Green's function of the isolated normal metal and superconductor take the form:
\begin{align}
    &\check g_N^0= \tau_z, \label{eq:isolated_normal} \\
    &\check g_S^0 = \frac{\omega \tau_z}{\sqrt{\omega^2+|\Delta|^2}}  - \frac{i\Delta \tau_+ - i\Delta^* \tau_-}{\sqrt{\omega^2+|\Delta|^2}},
\end{align}
where $\tau_\pm = (\tau_x +\pm i \tau_y)/2$.

\begin{figure}[t]
\includegraphics[width=76mm]{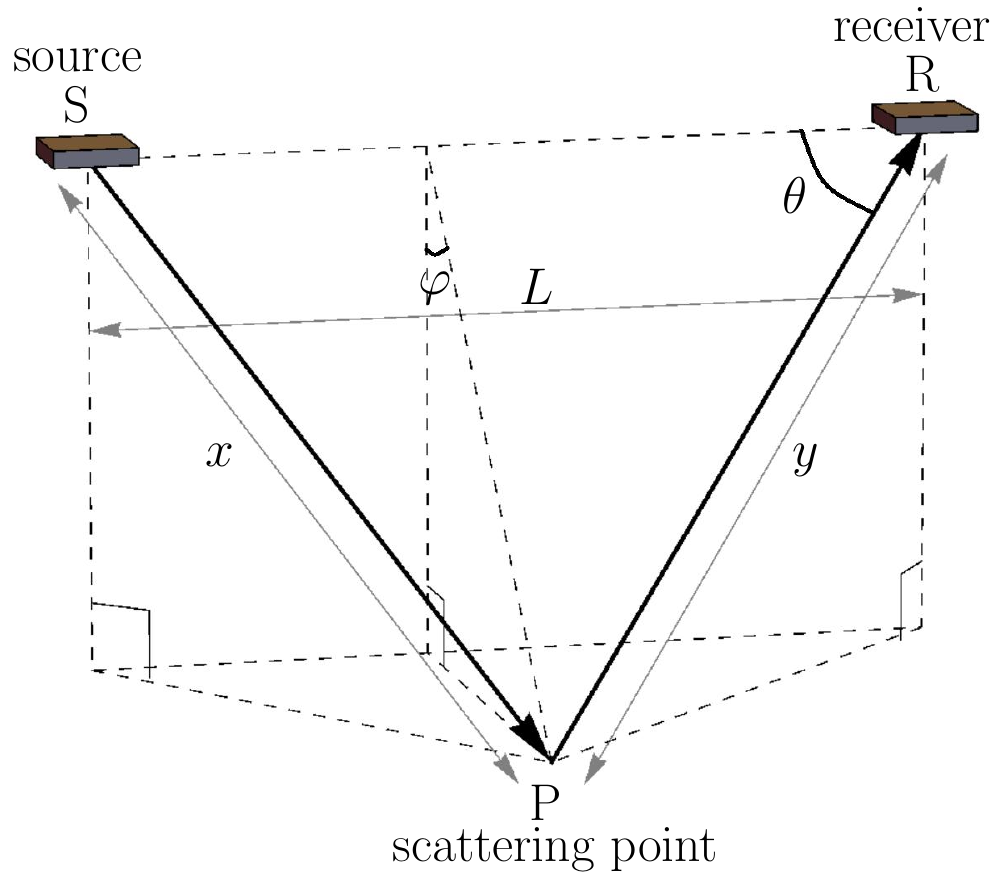}
\caption{Sketch of the trajectories between two small superconducting elements S and R contributing to the current flow between S and R via a single scattering event at the scattering point P.}
 \label{fig:elements}
\end{figure}

\subsection{Calculation of the Josephson supercurrent}

Regardless of the specific scattering model (from impurities or from the boundary), the condition $D \ll 1$ allows us to compute the Josephson current for an arbitrary configuration of superconducting leads. This holds both for a contact with just two superconducting leads and for any complex configuration involving multiple superconducting electrodes deposited on the surface of the normal crystal. To this end, we first need to calculate the current $I_{S\to R}$ between two small superconducting elements, which we denote as source (S) and receiver (R), as illustrated in Fig.~\ref{fig:elements}. Under the condition $D \ll 1$, the Green's function for each trajectory between any two elements can be treated independently of the presence or absence of other superconducting elements. Consequently, the principle of linear superposition applies, and the total current $I$ between a pair of superconducting leads is obtained by summing the currents between all pairs of small superconducting elements belonging to the leads:
\begin{align}
    I=\sum \limits_{S,R} I_{S\to R}.
    \label{eq:I_total}
\end{align}

\subsubsection{Bulk impurity scattering model}
\label{bulk_imp_model}

At first we calculate $I_{S\to R}$ in the framework of the model of the single impurity scattering in the bulk of the N layer, Fig.~\ref{fig:setup}(b) and then modify the calculation to calculate the Josephson current in the framework of two other models. 

Then one can construct a perturbation theory expansion based on the inverse scattering time: $\check g=\check g^0+\frac{1}{\tau} \check g^1+...$. We restrict ourselves by the first order term, what corresponds to our assumption of a single impurity scattering. Let's denote $\check g_{UW}$ the Green's function on the trajectory from $U$ to $W$ (it is called trajectory $UW$) for arbitrary points $U,W$. Let consider an arbitrary point $P$. Our goal is calculate the Green's function at SP/PS and PR/RP trajectories up to the first order with respect to $\tau^{-1}$. Let us denote for PR/RP trajectories the distance from point R to P as $y$ and for SP/PS trajectories the distance from point S to P as $x$ and $x=\sqrt{L^2+y^2-2Ly\cos\theta}$ where $L$ is a distance between points $S$ and $R$.

As a first step we find the anomalous Green's functions at SP/PS and PR/RP trajectories up to the zero order with respect to $\tau^{-1}$ from the Eilenberger equation (\ref{eq:eilenberger}) supplemented by boundary conditions at the superconductor/normal metal interfaces (\ref{eq:bc}) taken at S and R points. For the incoming trajectory PS the Eilenberger equations (\ref{eq:eilenberger}) for the anomalous Green's functions take the simple form:
\begin{align}
-v_F \frac{d f}{d x} + 2 \omega f + \frac{2}{\tau}(\langle g \rangle f - \langle f \rangle g)= 0. 
    \label{eq:incoming}
\end{align}
\begin{align}
-v_F \frac{d \tilde f}{d x} - 2 \omega \tilde f - \frac{2}{\tau}(\langle g \rangle \tilde f - \langle \tilde f \rangle g)= 0. 
    \label{eq:incoming2}
\end{align}
The same is valid for the incoming trajectory PR with the substitution $x \to y$. The solution of the Eilenberger equation for these trajectories at $\tau^{-1}=0$ takes the form $f_{PS}^0 = C_{PS}e^{\lambda x}$ and $f_{PR}^0 = C_{PR}e^{\lambda y}$, where $\lambda=\frac{2\omega}{v_F}$. Further we assume $\omega >0$. Then $C_{PS}=C_{PR}=0$ from the asymptotic condition $f,\tilde f \to 0$ at $x,y \to \infty$.  
For the outgoing trajectory SP the Eilenberger equations for the anomalous Green's functions are obtained from Eqs.~(\ref{eq:incoming})-(\ref{eq:incoming2}) by the substitution $v_F \to -v_F$.
The solution at $\tau^{-1}=0$ takes the form $f_{SP}^0 = C_{SP}e^{-\lambda x}$ and $f_{RP}^0 = C_{RP}e^{-\lambda y}$, where $\lambda=\frac{2\omega}{v_F}$. The asymptotic condition is fulfilled automatically for these trajectories. $C_{SP(RP)}$ are to be found from the boundary condition (\ref{eq:bc}), which for the anomalous Green's function takes the form: 
\begin{align}
&f^{out}(x(y)=0) = D f_{S(R)}^0 \nonumber \\
&\tilde f^{out}(x(y)=0) = -D \tilde f_{S(R)}^0,
    \label{eq:bc_anomalous}
\end{align}
where $f_{S(R)} = -i|\Delta|e^{i \phi_{1(2)}}/\sqrt{\omega^2 + |\Delta|^2}$ and $\tilde f_{S(R)} = i|\Delta|e^{-i \phi_{1(2)}}/\sqrt{\omega^2 + |\Delta|^2}$ with $\phi_{1(2)}$ being the phase of the superconducting order parameter of the S(R) element. Finally, the zero-order anomalous Green's functions for SP, PS, SR and RS trajectories take the form:
 \begin{align}
    f_{SP}^0=\frac{-iD|\Delta| e^{i\phi_1}}{\sqrt{|\Delta|^2+\omega^2}}e^{-\lambda x}, ~~~~\tilde f_{SP}^0=0 , \nonumber \\ 
    \tilde f_{PS}^0=\frac{iD|\Delta| e^{-i\phi_1}}{\sqrt{|\Delta|^2+\omega^2}}e^{-\lambda x}, ~~~~ f_{PS}^0=0, \nonumber \\ 
    \tilde f_{PR}^0=\frac{iD|\Delta| e^{-i\phi_2}}{\sqrt{|\Delta|^2+\omega^2}}e^{-\lambda y}, ~~~~ f_{PR}^0=0, \nonumber \\ 
    f_{RP}^0=\frac{-iD|\Delta| e^{i\phi_2}}{\sqrt{|\Delta|^2+\omega^2}}e^{-\lambda y}, ~~~~ \tilde f_{RP}^0=0 .
    \label{eq:anomalous_zero}
\end{align}
Then from the normalization condition it follows that $g_{UW}^0 = 1$ for all the trajectories and equation for the first order correction to $f_{PR}$ can be written as follows:
\begin{align}
    -\frac{2 }{v_F} \langle f^0 \rangle +\lambda f_{PR}^1-\frac{d}{dy} f_{PR}^1=0 .
    \label{eq:eq_first_PR}
\end{align}
The solution of this equation 
\begin{align}
    f_{PR}^1(y)=e^{\lambda y}\int\limits_y^\infty e^{-\lambda z} \frac{2}{v_F}\langle f^0(z) \rangle dz
    \label{eq:solution_first_PR}
\end{align}
contains the averaged over the Fermi surface value of the zero-order anomalous Green's function, which is to be calculated as 
\begin{align}
    \langle f^0(y) \rangle = \frac{d\Omega_S}{4\pi}f^0_{SP}+\frac{d\Omega_R}{4\pi}f^0_{RP} .
    \label{eq:f_average}
\end{align}
The solid angles of the source and receiver are expressed as
\begin{align}
    d\Omega_S=\frac{dS_S y \sin \theta \cos \varphi}{x^3},~~~
    d\Omega_R=\frac{dS_R \sin \theta \cos \varphi}{y^2} .
    \label{eq:solid_angles}
\end{align}
At $\omega < 0$ the solutions for the anomalous Green's function can be obtained from Eqs.~(\ref{eq:anomalous_zero}), (\ref{eq:solution_first_PR}) and (\ref{eq:f_average}) by interchanging the incoming and outgoing trajectories $SP(R) \leftrightarrow P(R)S$.

The electric current should be calculated via the normal Green's function $g$,
\begin{align}
    I=-i\pi N_F T\sum_\omega\int d\Omega \bm v_F g 
\end{align}
which can be expressed via the anomalous Green's function making use of the normalization condition. Up to the lowest order with respect to the anomalous Green's function it takes the form:
\begin{align}
g=1-f \tilde f/2 .   
    \label{eq:g}
\end{align}
Then 
\begin{align}
g_{PR}=1-f_{PR}^1 \tilde f_{PR}^0/2 , \nonumber \\
g_{RP}=1-f_{RP}^0 \tilde f_{RP}^1/2
    \label{eq:g_trajectory}
\end{align}
It is convenient to calculate the electric current between S and R at $y=0$:
\begin{align}
    I_{S\to R}=-i\pi N_F T \sum \limits_\omega \int\limits_0^\pi \int\limits_{-\pi/2}^{\pi/2} \delta I(\theta, \varphi) \sin\theta d\theta d\varphi, \nonumber \\
     \delta I(\theta, \varphi)=v_F [g_{PR}(\theta,\varphi,y=0)- g_{RP}(\theta,\varphi,y=0)] ,
     \label{eq:current_element_general}
\end{align}
where $N_F$ is the density of states at the Fermi level summed over spin. Substituting the normal Green's functions into Eq.~(\ref{eq:current_element_general}) we obtain the following expression for $I_{S \to R}$ for single scattering event at an impurity in the bulk of the N crystal:
\begin{align}
    I_{S\to R}^a=T \sum \limits_\omega \frac{\pi eD^2|\Delta|^2 dS_S dS_R N_F \lambda \sin(\phi_2-\phi_1)}{\tau (|\Delta|^2+\omega^2)} \times ~~~~~~~~ \nonumber \\
    \int\limits_0^\pi\int \limits_0^\infty e^{-\tilde z-\sqrt{\tilde z^2+\tilde L^2-2\tilde L \tilde z\cos \theta}}\frac{\tilde z \sin^2\theta d\theta d\tilde z}{(\tilde L^2+\tilde z^2-2\tilde L \tilde z \cos\theta)^{3/2}}
    \label{eq:current_a}
\end{align}
where $\tilde L=\lambda L$. A very good approximation for the integral in the second line of Eq.~(\ref{eq:current_a}) is
\begin{align}
1.3 e^{-1.21 \lambda L}+1.7 e^{-3.3 \lambda L},
\end{align}
and the approximate expression for the total current between the S and R elements takes the form:
\begin{align}
    I_{S\to R}^a \approx &T \sum \limits_\omega \frac{\pi eD^2|\Delta|^2 dS_S dS_R N_F \lambda \sin(\phi_2-\phi_1)}{\tau (|\Delta|^2+\omega^2)} \times ~~~~~~~~ \nonumber \\
   & (1.3 e^{-1.21 \lambda L}+1.7 e^{-3.3 \lambda L}).
    \label{eq:current_a_approximate}
\end{align}
The accuracy of this expression and of the analogous approximations obtained below for a thin normal crystal, Eqs.~\ref{eq:current_element_specular_approximate}) and (\ref{eq:current_diffusive_surface_approximate}), was estimated over the parameter range \( \lambda L \) from 0 to \(\sim 5\). This range spans from the regime where source and receiver are very close to each other up to separations exceeding the coherence length manyfold. In those expressions where the crystal thickness \(d\) appears, we also examined \(d/L\) from 0 to \(\sim 0.5\). Over this entire parameter domain the deviation of the proposed formulas from the exact numerical evaluation remains within approximately \(5\%\).

The motivation for introducing these approximations is twofold. First, the exact expressions are computationally demanding; their direct use in all subsequent calculations would be technically prohibitive. Second, the approximate formulas render the functional dependence on the key parameters transparent, in contrast to the exact integral or sum representations, thereby providing physical insight into the behavior of the system.

\subsubsection{Specular reflection model for a thin normal crystal}

\begin{figure}[t]
\includegraphics[width=84mm]{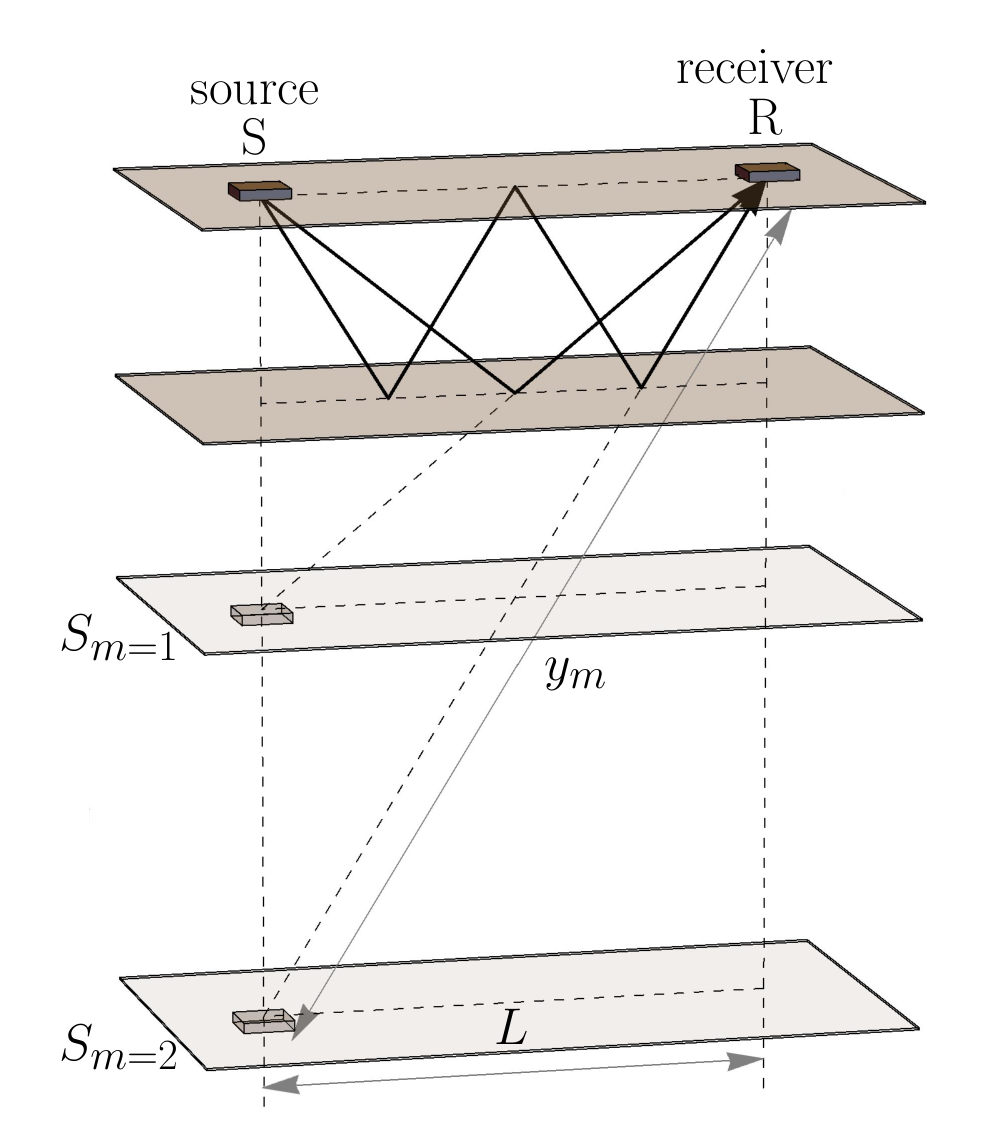}
\caption{Sketch of the trajectories between two small superconducting elements S and R contribution to the current flow between S and R via a multiple mirror reflections.}
 \label{fig:mirror_reflection}
\end{figure}

Now let us consider the case of the thin normal crystal with $d \ll (\xi,l)$ with plane surfaces without roughness. In this case a quasiparticle trajectory between S and R elements exhibits multiple specular reflections at the  upper and lower surfaces of the normal layer, as it is sketched in Fig.~\ref{fig:setup}(c). Then the boundary condition at the N upper and lower surfaces takes the form:
\begin{align}
    \check g^{out} = \check g^{in} .
    \label{eq:specular}
\end{align} 
Now $\check g^{in(out)}$ is the Green's function incoming to (outgoing from) the surfaces of the N layer. In this case one can use the method of image sources, see Fig.~\ref{fig:mirror_reflection}, and replace the trajectory containing the $m$-fold reflection from the bottom surface by the straight trajectory between the image $\rm S_m$ and R elements. For $m$-fold reflections from the bottom surface (the corresponding position $\rm S_m$ is located under the source at a depth of $2md$) the solid angles of the image source $\rm S_m$ takes the form:
\begin{align}
d\Omega_{S_m}=dS_S\frac{2md}{(L^2+(2md)^2)^{3/2}}.  
\end{align}
The anomalous Green's functions can be found from Eq:~(\ref{eq:eilenberger}) with $\tau^{-1} = 0$. At the R point corresponding to $y=0$ (taking into account distance between R and $\rm S_m$ points $y_m = \sqrt{L^2+(2md)^2}$) the anomalous Green's functions take the form:
\begin{align}
&f_{m, SR}=\frac{-iD|\Delta|e^{i\phi_1}}{\sqrt{|\Delta|^2+\omega^2}}e^{-\lambda\sqrt{L^2+(2md)^2}}, \nonumber \\
&\tilde f_{m, SR} = \frac{iD|\Delta|e^{-i\phi_2}}{\sqrt{|\Delta|^2+\omega^2}}, \nonumber \\
&f_{m, RS} = \frac{-iD|\Delta|e^{i\phi_2}}{\sqrt{|\Delta|^2+\omega^2}}, \nonumber \\
&\tilde f_{m, RS}=\frac{iD|\Delta|e^{-i\phi_1}}{\sqrt{|\Delta|^2+\omega^2}}e^{-\lambda\sqrt{L^2+(2md)^2}} .
\label{eq:anomalous_specular}
\end{align}
The net electric current between S and R can be calculated making use of Eqs.~(\ref{eq:g}) and (\ref{eq:current_element_general}) at $y=\sqrt{L^2+(2md)^2}$ and takes the form
\begin{align}
I_{S\to R}^b=&T \sum \limits_\omega \frac{\pi eD^2|\Delta|^2 dS_S dS_R N_F v_F\sin(\phi_2-\phi_1)}{ |\Delta|^2+\omega^2} \times  \nonumber \\
&\sum\limits_{m=0}^{\infty} \frac{2md}{(L^2+(2md)^2)^{3/2}}e^{-\lambda\sqrt{L^2+(2md)^2}} ,
\label{eq:current_element_specular}
\end{align}
which can be approximated to a rather good accuracy by 
\begin{align}
I_{S\to R}^b=&T \sum \limits_\omega \frac{\pi eD^2|\Delta|^2 dS_S dS_R N_F v_F\sin(\phi_2-\phi_1)}{|\Delta|^2+\omega^2} \times  \nonumber \\
&\frac{\lambda}{2d}\left( \frac{e^{-\lambda\sqrt{L^2+4d^2}}}{\lambda\sqrt{L^2+1.5d^2}}+{\rm Ei}(-\lambda \sqrt{L^2+4d^2}) \right).
\label{eq:current_element_specular_approximate}
\end{align}
This model gives us a generalization of expressions found in \onlinecite{Kuprianov1986,Kuprianov1988} for the case of an arbitrary geometry of the superconducting leads.

\subsubsection{Diffusive reflection model for a thin normal crystal}
\label{film_diffusive_model}

Further we consider the thin normal crystal with $d \ll (\xi, l)$ and rough surfaces. In the case of the isotropic diffusive scattering and to the first order with respect to $D \ll 1$ the boundary condition at the surfaces takes the form \cite{Eschrig2009}:
\begin{align}
    f^{out} = \langle f \rangle_{in} , ~~ \tilde f^{in} = \langle \tilde f \rangle_{out} ,
    \label{eq:diffusive}
\end{align}
where $\langle  \rangle_{in(out)}$ means averaging over all the trajectories incoming to (outgoing from) the bottom surface of the N layer. We take into account only single scattering event from the bottom surface, and the multiple diffusive reflections from the upper and bottom surfaces are disregarded, see Fig.~\ref{fig:setup}(d). Let the scattering point at the bottom surface of the N layer is P. The distance between R and P is $|PR|$. Then from Eq.~(\ref{eq:incoming}) with $\tau^{-1} = 0$ and taking into account boundary condition (\ref{eq:diffusive}) we obtain the following expression for the anomalous Green's function at the PR trajectory (the notations are the same as in the case of a single scattering at the bulk impurity):
\begin{align}
    f_{PR}(y)=\langle f^0(y) \rangle_{in} e^{-\lambda (|{\rm PR}|-y)},
    \label{eq:anomalous_surface_diffusive}
\end{align}
where 
\begin{align}
    y=\frac{d}{\cos \varphi \sin \theta},~~~x=\sqrt{L^2+y^2-2Ly\cos \theta},
\end{align}
\begin{align}
    \langle f^0(y) \rangle_{in} = \frac{d\Omega_S}{2\pi}f^0_{SP}+\frac{d\Omega_R}{2\pi}f^0_{RP} 
    \label{eq:f_average_surface}
\end{align}
and $f_{SP}^0$ and $f_{RP}$ are expressed by Eqs.~(\ref{eq:anomalous_zero}). Calculating the anomalous function $\tilde f_{RP}$ in the same way and expressing the normal Green's function via the anomalous one as
\begin{align}
g_{PR}=1-f_{PR} \tilde f_{PR}^0/2 , \nonumber \\
g_{RP}=1-f_{RP}^0 \tilde f_{RP}/2,
    \label{eq:g_trajectory_surface}
\end{align}
from Eq.~(\ref{eq:current_element_general}) we obtain the following expression for the electric current between S and R superconducting elements:
\begin{align}
    I_{S\to R}^c=&T \sum \limits_\omega \frac{D^2e|\Delta|^2 dS_S dS_R N_F \lambda^2 v_F\sin(\phi_2-\phi_1)}{2 (|\Delta|^2+\omega^2)} \times \nonumber \\ 
    &\int\limits_0^\pi\int \limits_{-\pi/2}^{\pi/2} \frac{e^{\tilde d-\sqrt{\tilde d^2+\tilde L^2-2\tilde L\tilde d \cos\theta}}\tilde d}{(\tilde d^2+\tilde L^2-2\tilde L\tilde d)^{3/2}} \sin \theta d\theta d\varphi ,
\label{eq:current_diffusive_surface}
\end{align}
where $\tilde L=\lambda L, \tilde d=\frac{\lambda d}{\cos \varphi \sin \theta}$. Rather good approximation for the integral in the second line of this expression is
\begin{align}
\frac{2}{\lambda^2(L^2+4d^2)}(2e^{-\lambda\sqrt{L^2+4d^2}}+e^{-\lambda (L+2d)}).
\end{align}
Then the approximate expression for the the electric current between S and R superconducting elements in the considered model is 
\begin{align}
    I_{S\to R}^c=&T \sum \limits_\omega \frac{D^2e|\Delta|^2 dS_S dS_R N_F \lambda^2 v_F\sin(\phi_2-\phi_1)}{2 (|\Delta|^2+\omega^2)} \times \nonumber \\ 
    & \frac{2}{\lambda^2(L^2+4d^2)}(2e^{-\lambda\sqrt{L^2+4d^2}}+e^{-\lambda (L+2d)}).
\label{eq:current_diffusive_surface_approximate}
\end{align}
If one takes into account not only single, but also multiple scattering events from the N layer surfaces, then the current should contain additional contributions with reduced exponential factors $\exp[-\lambda \sqrt{L^2 + (2 \kappa d)^2}]$ or $\exp[-\lambda (L+ 2 \kappa d)]$, where $\kappa >1$ and grows with increasing the number of diffusive reflection from the N surfaces. This means that the single scattering approximation works if $\lambda d \gg1$ and, consequently $d/\xi_T \gg 1$, where $\xi_T = v_F/2 \pi T$ is the coherence length at a given temperature. The other reducing factor in the contribution of multiple scattering events is $1/[L^2 + (2 \kappa d)^2]$. This means that multiple scattering events can be neglected also in the case $L<2d$.

\subsubsection{Effect of an external magnetic field on the supercurrent}

In the presence of an external magnetic field, for all cases (impurity/boundary scattering) the expressions for the electric current should be generalized  in a standard way \cite{Barzykin1999} by adding the integral of the magnetic field vector potential $\bm A$ to the superconducting phase difference:
\begin{align}
    (\phi_2-\phi_1)_A =  \phi_2-\phi_1+\frac{2 \pi}{\Phi_0}\int \bm A(s) \cdot \bm n ds
\end{align}
where $\bm n$ is the unit vector along the trajectory and $s$ is the coordinate along it. The important point is that now the gradient invariant phase difference depends on the quasiparticle trajectory. For this reason the expression for the electric current in the models of the single diffusive scattering from the bulk impurity and from the single diffusive scattering from the surface should be rewritten as follows:
\begin{align}
    I_{S\to R}^a=T \sum \limits_\omega \frac{eD^2|\Delta|^2 dS_S dS_R N_F \lambda }{2\tau (|\Delta|^2+\omega^2)} \times ~~~~~~~~ \nonumber \\
    \int\limits_0^\pi\int \limits_0^\infty e^{-\tilde z-\sqrt{\tilde z^2+\tilde L^2-2\tilde L \tilde z\cos \theta}}\frac{\sin(\phi_2-\phi_1)_A \tilde z \sin^2\theta d\theta d\tilde z}{(\tilde L^2+\tilde z^2-2\tilde L \tilde z \cos\theta)^{3/2}}
    \label{eq:current_a_field}
\end{align}
and
\begin{align}
    I_{S\to R}^c=T \sum \limits_\omega \frac{D^2e|\Delta|^2 dS_S dS_R N_F \lambda^2 v_F\sin(\phi_2-\phi_1)}{8\pi (|\Delta|^2+\omega^2)} \times \nonumber \\ 
    \int\limits_0^\pi\int \limits_{-\pi/2}^{\pi/2} \frac{e^{\tilde d-\sqrt{\tilde d^2+\tilde L^2-2\tilde L\tilde d \cos\theta}}\tilde d}{(\tilde d^2+\tilde L^2-2\tilde L\tilde d)^{3/2}} \sin(\phi_2-\phi_1)_A \sin \theta d\theta d\varphi .
\label{eq:current_diffusive_surface}
\end{align}
The approximate expressions (\ref{eq:current_a_approximate}) and (\ref{eq:current_diffusive_surface_approximate}) are no longer valid. At the same time if the magnetic field is applied perpendicular to the surface of the normal crystal the electric current $I_{S \to R}^b$ calculated for the model of multiple specular reflections from the N layer surfaces is still expressed by Eqs.~(\ref{eq:current_element_specular}) and (\ref{eq:current_element_specular_approximate}) with the substitution $\phi_2 - \phi_1 \to (\phi_2 - \phi_1)_A$. The reason is that for the perpendicular magnetic field the vector potential $\bm A$ is directed in plane of the surface and the projection of each trajectory on the surface plane is the same straight line a connecting S and R elements.

\section{Josephson current of a single junction}
\label{current}

\begin{figure}[t]
\includegraphics[width=84mm]{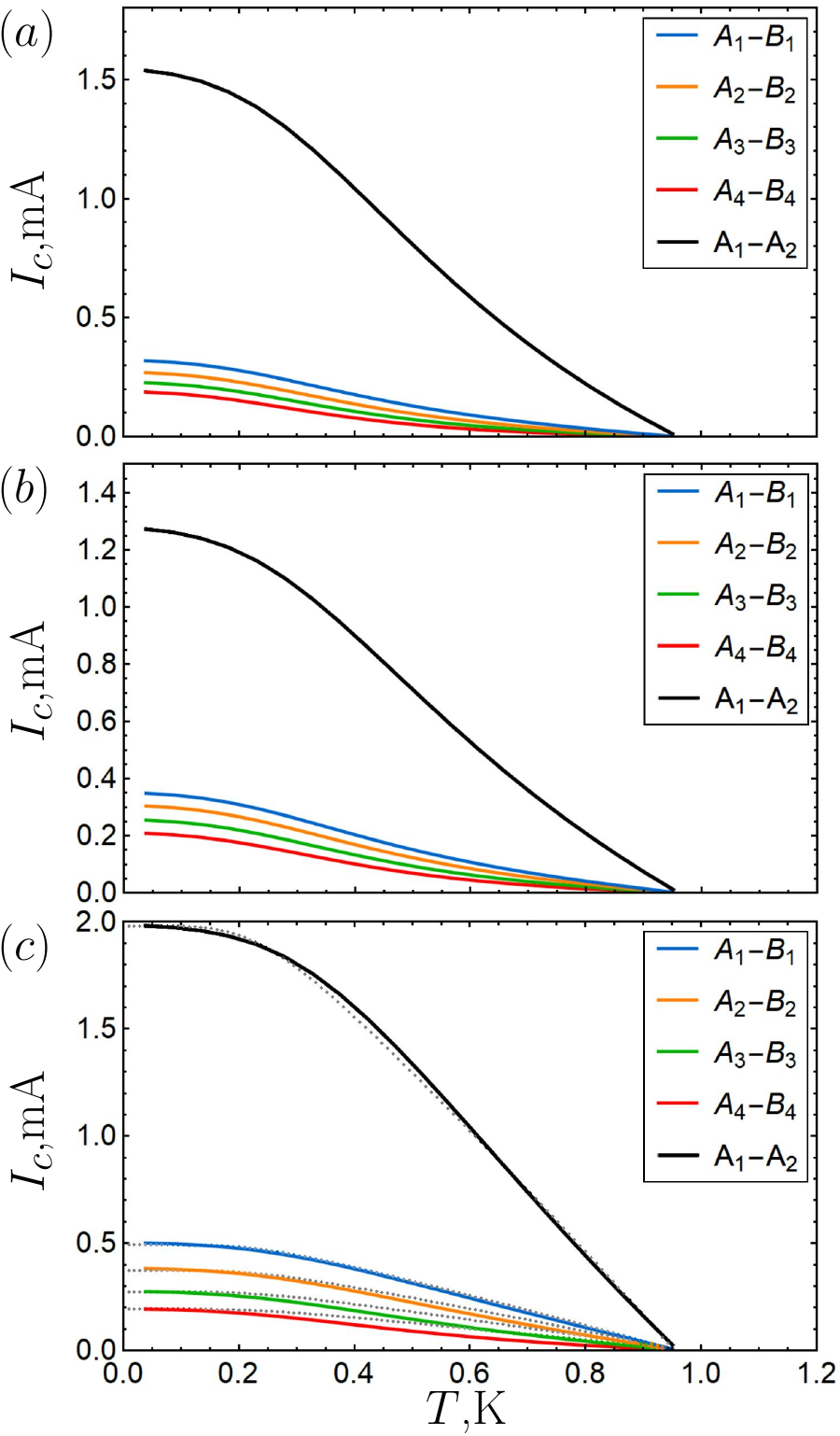}
\caption{Dependence of the critical Josephson current on temperature calculated for a JJ between two superconducting leads in the absence of other leads on top of the N crystal. Different curves correspond to different JJs between the leads marked by the same letters in Fig.~\ref{fig:setup}. (a) The model of a single scattering at a bulk impurity. $D=0.004$. (b) The model of multiple specular reflections from the surfaces of the N crystal. $D=0.013$. (c) The model of a single diffusive reflection from the bottom boundary of the N layer. $D=0.03$. The dotted grey lines in panel (c) represent he Kulik-Omelyanchuk result for the junction with the same zero-temperature critical current. The transparency for each model is chosen to fit the experimental value of the multiterminal critical current \cite{Polevoy2025_joint} as described in the main text.}
 \label{fig:one_temperature}
\end{figure}

Now making use of the theoretical approach developed in the previous section we calculate the critical Josephson current between two superconducting leads in the framework of all three considered models. The results for the dependence of the corresponding critical current on temperature are shown in Fig.~\ref{fig:one_temperature}. Panels (a)-(c) of Fig.~\ref{fig:one_temperature} represent $I_c(T)$ calculated in the framework of the single scattering at a bulk impurity model (a), mupltiple specular reflections model (b) and single diffusive reflection from the bottom surface model (c). Different curves correspond to different geometry of the junctions. To be specific we chose geometrical parameters of the junctions to model the experimental junctions \cite{Polevoy2025_joint} $A_i-B_i$ and $A_1-A_2$ [see Fig.~\ref{fig:setup}(a)], but it is important that here in order to investigate the Josephson current step by step we consider the studied junction {\it in the absence of any other superconducting leads on top of the N crystal}. In calculating these temperature dependencies, we employed the well-known approximate expression for the temperature dependence of the superconducting order parameter $\Delta(T)=1.76\sqrt{1-(T/T_c)^3}$. The order parameter of the superconducting electrodes was not calculated self-consistently because the low transparency of the S/N interface allows us to neglect the inverse proximity effect, i.e., the influence of the normal layer on the superconductors. The complex Josephson structure with many leads is considered in the next section.

The general observation, which can be done by inspecting Fig.~\ref{fig:one_temperature} is that the temperature dependence of $I_c$ can be of two types. The first one is when with decreasing temperature the linear growth of $I_c$ becomes saturated, see  Fig.~\ref{fig:one_temperature}(c). The second type of behavior is when with decreasing temperature the initial linear growth of current becomes faster than linear and only then reaches saturation, see Figs.~\ref{fig:one_temperature}(a)-(b). The first type of behavior is similar to the Kulik-Omelyanchuk result generalized for low-transparency JJs \cite{Haberkorn1978} $I_c \propto |\Delta| \tanh[|\Delta|/2T]$. It is not surprising because the interlayer length of studied junctions $L_i < \xi$ and, roughly speaking, they meet the short junction limit. The faster than linear growth of $I_c$ with decreasing temperature is a consequence of the planar geometry. The point is that with decreasing temperature the temperature coherence length $\xi_T$ increases and as a consequence, an ever larger part of the superconducting leads (of the order of the coherence length from the interlayer region) begins to contribute to the Josephson current. Therefore, the current begins to grow compared to the case when the area of the superconducting leads does not depend on temperature. Different models give different power dependence of critical current on distance between source and receiver, see Eqs.~(\ref{eq:current_a_approximate}), (\ref{eq:current_element_specular_approximate}) and (\ref{eq:current_diffusive_surface_approximate}). In particular, in the model of single diffusive scattering from the boundary, the contribution of superconducting elements nearest to the interlayer increases compared to other models due to the inverse square dependence on distance between the elements. Therefore, this model is the least sensitive to effective increase of length of superconducting leads, contributing to current with decreasing temperature. This leads to the fact that this model gives the smallest deviation from linear behavior with decreasing temperature.

It is important to note that not only the choice of a specific model, but also the geometry of the superconducting leads strongly affects the final form of the temperature dependence of the Josephson current. In particular, within the framework of any of the considered models, the excess growth of the current with decreasing temperature disappears if the geometric dimensions of the superconducting leads become $\lesssim \xi$.

It is also worth noting that achieving the same amplitude of the critical current in the different models considered requires different values of the interface transparency $D$, see caption to Fig.~\ref{fig:one_temperature}. Importantly, within our approach, the transparency enters the current only as an overall amplitude factor, see Eqs.~(\ref{eq:I_total}), (\ref{eq:current_a_approximate}), (\ref{eq:current_element_specular_approximate}) and (\ref{eq:current_diffusive_surface_approximate}). Consequently, it does not serve as a fitting parameter for the temperature or field dependence of the critical current in the multiterminal system; instead, it merely renormalizes the overall current amplitude. Thus, for each model, the transparency was adjusted so that the multiterminal current $I_c^{A_i-B_i}(T=0,B=0)$ for an arbitrarily chosen pair of electrodes $A_i$ and $B_i$ matched the experimental value \cite{Polevoy2025_joint}. With this choice, all our models reproduce the ratios between different $I_c^{A_i-B_i}(T=0,B=0)$ and $I_c^{A_i-A_j}(T=0,B=0)$ without any fitting parameters. The single-junction currents shown in Fig.~\ref{fig:one_temperature} were obtained after computing the total current of the multiterminal system and fixing the transparency for each model in the manner described above.

\begin{figure}[t]
\includegraphics[width=84mm]{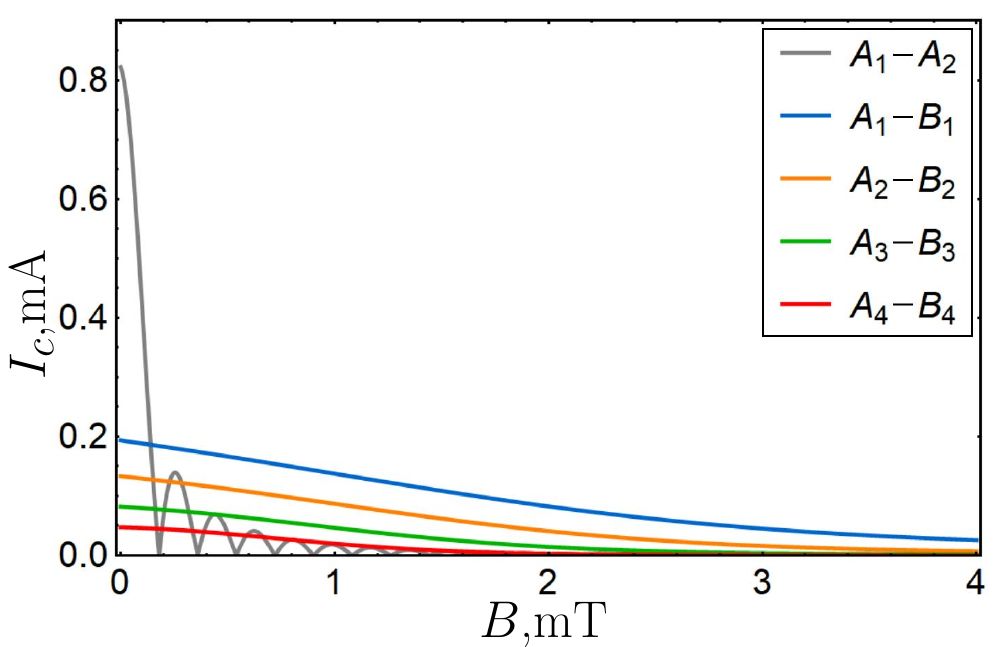}
\caption{Dependence of the critical Josephson current on the applied perpendicular magnetic field. Different curves correspond to the same JJs as in Fig.~\ref{fig:one_temperature}. The results are calculated in the framework of the model of a single diffusive reflection from the bottom N layer surface. $T=0.7K$.}
 \label{fig:one_field}
\end{figure}

Now we discuss the dependence of the critical superconducting current between two superconducting leads on the applied magnetic field, calculated in the absence of any other leads on top of the crystal. The field is applied perpendicular to the N crystal surface. The dependencies $I_c(B)$ are shown in Fig.~\ref{fig:one_field} for the same JJs that were discussed above in the context of the temperature dependence of the Josephson current. The current is calculated in the framework of a single diffusive reflection from the bottom surface of the N layer. The other two models give very similar results. $I_c(B)$ strongly depends on the geometry of the JJ and  varies from the standard Fraunhofer picture for wide junction with $w \gg \xi$ (where $w$ is the width of the interlayer region) to a monotonic decrease of $I_c$ with increasing field for narrow junctions with $w < \xi$. Similar crossover is typical for diffusive sandwich S/N/S Josephson junctions \cite{Cuevas2007} and differs from ballistic non-planar JJs, where the dependence on the magnetic field manifests magnetic interference patterns and has a periodic or quasi-periodic character depending on the geometry of the JJ \cite{Barzykin1999,Ledermann1999}. In the case under consideration, the absence of the magnetic interference pattern for narrow JJs is a consequence of the smearing of a clear picture of closed-loop current-carrying Andeev bound states in the case of planar junction geometry.

\section{Complex supercurrent flow in multi-terminal Josephson systems}
\label{multiterminal}

\begin{figure}[t]
\includegraphics[width=84mm]{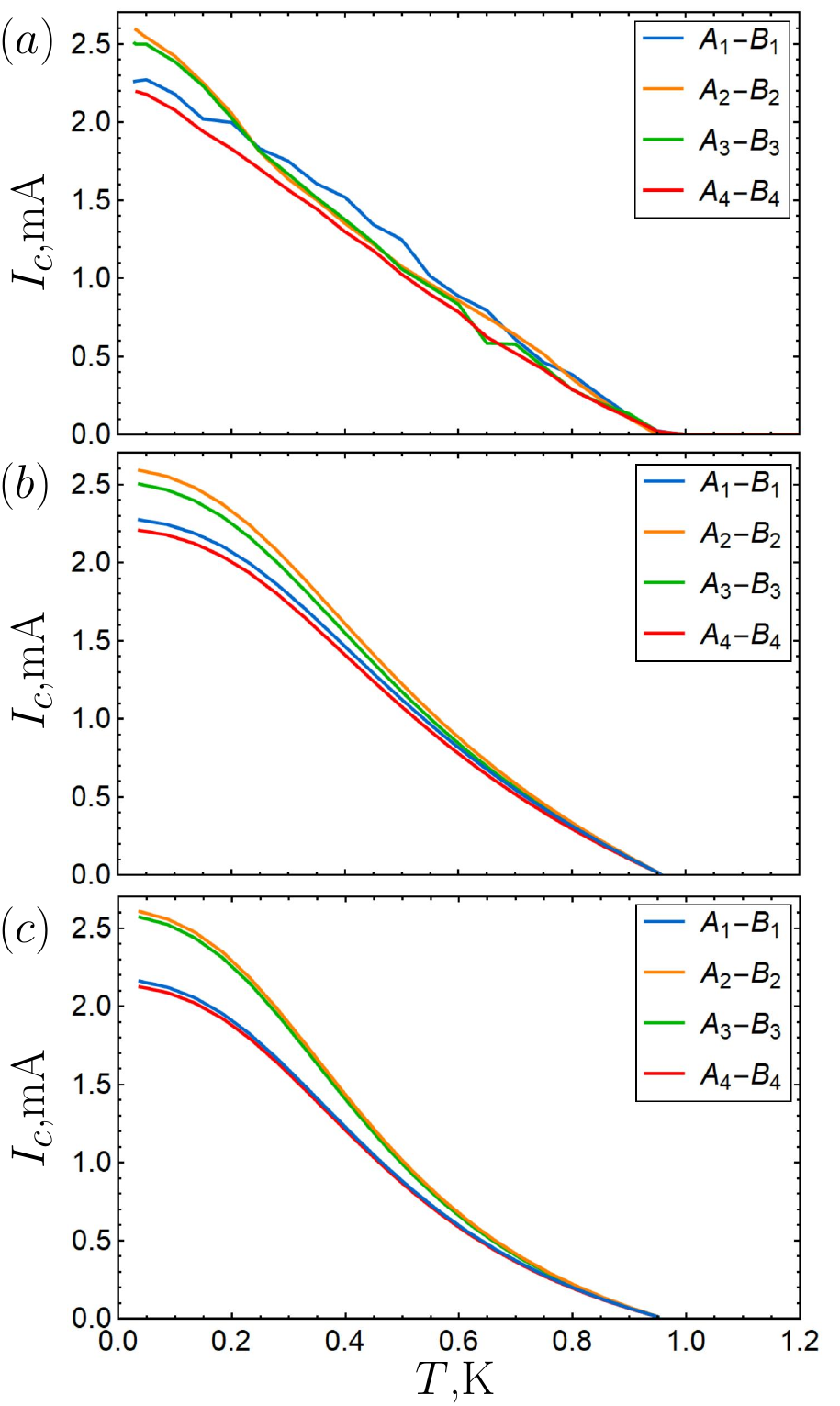}
\caption{(a) Experimental data \cite{Polevoy2025_joint} for the critical current $I_c^{A_i-B_i}$ vs temperature in the multi-terminal setup, shown in Fig.~\ref{fig:setup}. (b) Theoretical results for $I_c^{A_i-B_i}$ vs temperature in the same multi-terminal setup in the framework the model of a single diffusive reflection from the bottom N layer surface and (c) in the framework of the model of multiple specular reflections from the surfaces of the N layer. }
 \label{fig:multi_temperature}
\end{figure}

The key experimental observation \cite{Polevoy2025_joint} is that in the complex planar setup shown in Fig.~\ref{fig:setup} the dependence of $I_c$ on the interlayer length $L_{i}$ is nonmonotonic, see Fig.~\ref{fig:multi_temperature}(a). The critical currents of two middle JJs $I_c^{A_2-B_2}(T=200 \mathrm {mK}) \approx I_c^{A_3-B_3}(T=200 \mathrm {mK}) = I_c^m$  are very close to each other and exceed the currents of the two outer JJs $I_c^{A_1-B_1}(T=200 \mathrm {mK}) \approx I_c^{A_4-B_4}(T=200 \mathrm {mK}) = I_c^o$, which are also close to each other and by a factor of $I_c^o/I_c^m \approx 7/8$ smaller than $I_c^m$. This means that the 4 experimentally investigated JJs are not independent. Indeed, taking the values of the Fermi velocity $v_F=1.4\cdot 10^{6}m/s$ for Au and the critical superconducting temperature $T_c = 0.95{\rm K}$ of Al electrodes, the ballistic coherence length for the Au sample can be estimated as $\xi = v_F/2\pi T_c = 1.7\mu$m, which significantly exceeds the $L_{i}$ for all the $A_i-B_i$ JJs. Also, this value exceeds or is at least comparable with the distances between all superconducting leads $L_{A_i-A_j}$ and  $L_{B_i-B_j}$. Thus, in this experiment we are dealing with a complex multi-terminal Josephson system in which the Josephson current measured between any two superconducting leads is not equivalent to the Josephson current calculated in the previous section in the absence of other superconducting leads on top of the N crystal. 

\begin{figure}[t]
\includegraphics[width=84mm]{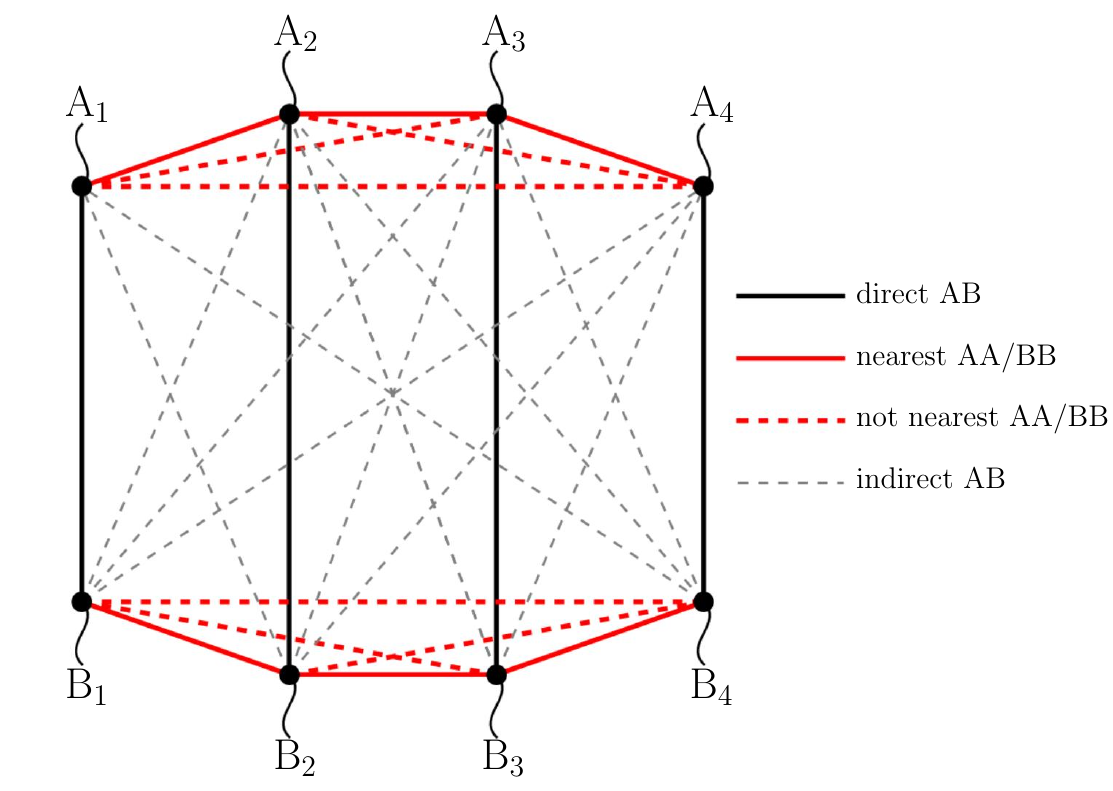}
\caption{An equivalent circuit diagram that represents the
coupling between each pair of superconducting leads. The leads are shown by black points and letters $A_i$, $B_i$. Each JJ between a pair of the leads is depicted by a line. The system contains 28 JJs in total.}
 \label{fig:eq_circuit}
\end{figure}

The equivalent circuit diagram of this complex Josephson structure in presented in Fig.~\ref{fig:eq_circuit}. The system contains 28 JJs in total. To calculate the net critical current  of the fully connected Josephson junction array $I_c^{i_{\rm in}-i_{\rm out}}$ (where $i_{\rm in},i_{\rm out}\in \Omega=\{A_{1-4},B_{1-4}\}$) corresponding to the current flow from $i_{\rm in}$ to $i_{\rm out}$ electrode, we first calculate the currents $i_c^{A(B)_i-A(B)_j}$ between all possible pairs of $A(B)_i-A(B)_j$ and $A_i-B_j$ as it is described in the previous section. Then the superconducting phases $\varphi_{i_{\rm in}},\varphi_{i_{\rm out}}$ are set at two leads through which the current is passed and other phases are determined numerically from the condition of the current conservation. The detailed numerical  procedure is as follows. On all leads except the input and output leads $i_{\rm in},i_{\rm out}$, the current divergence vanishes, i.e., Kirchhoff’s current law holds:
\begin{align}
    \forall i\neq i_{\rm in},i_{\rm out}:~~~\sum_{j\in\Omega \setminus i} i_c^{i-j} \sin(\varphi_i-\varphi_j)=0.
\end{align}
At the input and output leads, however, the divergence equals the externally passing current $I^{i_{\rm in}-i_{\rm out}}$.

The phases  at the input and output leads $\varphi_{i_{\rm in}},\varphi_{i_{\rm out}}$ are known and fixed, while $I^{i_{\rm in}-i_{\rm out}}$ is not known a priori. The goal is therefore to determine the phase distribution $\varphi_i$ such that the zero-divergence condition is satisfied at all internal leads.

This is achieved numerically via an iterative procedure, starting from an arbitrary initial phase configuration $\varphi_i$ for $i\neq i_{\rm in},i_{\rm out}$. At each iteration, the phase at every internal lead is updated by an amplitude proportional to the local current imbalance:
\begin{align}
    \varphi_i \to \varphi_i + \delta \sum_{j\in\Omega \setminus i} i_c^{i-j} \sin(\varphi_i-\varphi_j).
\end{align}
The process converges and is continued until charge conservation is fulfilled at all internal leads. The resulting value of $I^{i_{\rm in}-i_{\rm out}} = \sum_{j\in\Omega \setminus i_{\rm in}} i_c^{i_{\rm in}-j} \sin(\varphi_{i_{\rm in}}-\varphi_j)= \\-\sum_{j\in\Omega \setminus i_{\rm out}} i_c^{i_{\rm out}-j} \sin(\varphi_{i_{\rm out}}-\varphi_j)$ at the external contacts then gives the current through the system for the specified phase difference $\varphi_{i_{\rm in}}-\varphi_{i_{\rm out}}$.

This procedure is repeated for a full range of phase differences, and the critical current $I^{i_{\rm in}-i_{\rm out}}_c$ is identified as the maximum value of the resulting current.

The calculated $I_c^{A_i-B_i}$ JJs for $i=1-4$ are presented in Fig.~\ref{fig:multi_temperature}(b)-(c). The calculations are performed in the framework of the models of a single diffusive reflection from the bottom N layer surface [Fig.~\ref{fig:multi_temperature}(b)] and multiple specular reflections from the N layer surfaces [Fig.~\ref{fig:multi_temperature}(c)]. The model of a single scattering from a bulk impurity is not relevant to the experimental data presented in Fig.~\ref{fig:multi_temperature}(a) because the thickness $d = 80$ nm of the Au crystal in the experimental setup \cite{Polevoy2025_joint} was much less than the mean free path $l=680$nm in the Au crystal. Both models provide very good quantitative agreement for the low-temperature values of all four critical currents $I_c^{A_i-B_i}$. The only fitting parameter is the transparency of the S/N interface $D$. However, it only controls the absolute value of the critical current amplitude, while the ratio between different $I_c^{A_i-B_i}$ is reproduced without any fitting parameters as a result of taking into account the presence of several JJs providing the complex current flow. The specific value of the ratio $I_c^o/I_c^m \approx 7/8$ is determined by  details of the specific experimental implementation. One of the main factors, which influence the value of the ratio is the specific value of $i_c^{A_i-A_j}$. If $i_c^{A_i-A_j}$ were much larger than $i_c^{A_i-B_i}$, the electric current through any pair of opposite superconducting leads $A_i$ and $B_i$ would be the same and equal to $I_c^{A_i-B_i}=\sum \limits_{i=1}^4 i_c^{A_i-B_i}$. The difference between $I_c^{A_i-B_i}$ appears accounting for finite values of the ratio $i_c^{A_i-A_j}/i_c^{A_i-B_i} \neq \infty$, and the critical currents $I_c^{A_i-B_i}$ tend to the corresponding values of the individual JJs $i_c^{A_i-B_i}$ if $i_c^{A_i-A_j} \to 0$ and $i_c^{A_i-B_j} \to 0$.

The overall nearly linear temperature dependence of the critical current is better reproduced by the model of a single diffusive reflection [Fig.~\ref{fig:multi_temperature}(b)], although the quantitative agreement between theory and experiment at intermediate temperatures is not ideal. In the experiment, $I_c$ begins to grow with decreasing temperature faster than the linear dependence at lower temperatures than it is given by theory. In our opinion, this is due to the fact that in the experiment interlayer lengths of the most part of JJs are of the same order as the mean free path $l$ and, therefore, the ballistic limit condition is not fully satisfied, i.e. in addition to diffusive reflection from the surfaces of the N layer, there is also a certain amount of reflection from impurities in the N layer itself, which changes the temperature dependence of the coherence length. 

The response of a multi-terminal Josephson setup on the applied perpendicular magnetic field strongly depends on the particular geometry of the setup and can be very different and unusual due to the interplay of very different $i_c(B)$ for different individual JJs. The particular dependence $I_c^{A_i-B_i}(B)$ for the experimental configuration studied in Ref.~\onlinecite{Polevoy2025_joint} is discussed in the companion paper\cite{Polevoy2025_joint}. The most common characteristic feature of such multi-terminal systems is the superposition of monotonic dependences of narrow JJs and the Fraunhofer magnetic interference patterns of wide JJs, which are quasi-periodic due to slightly different geometry of the JJs.

\section{Conclusions}
\label{conclusions}

A theoretical framework based on the quasiclassical Eilenberger equations to describe quasi-ballistic planar Josephson junctions and strongly coupled complex structures of such junctions fabricated on the surface of a large crystal is developed. Three different models for calculation of the Josephson current between superconducting leads of an arbitrary geometry on top of a large quasi-ballistic normal metal crystal are considered. The first model is relevant to thick normal crystals and assumes that electrical transport between superconducting leads occurs through a single scattering process at an impurity in the bulk of the crystal. The second and third models are relevant to thin-film normal crystals with thicknesses smaller than the mean free path. The second model is applicable in the case of plane surfaces of a normal metal film without roughness and assumes that the current is transferred through multiple specular reflections from its surfaces. The third model applies to the case of diffuse scattering at the film surfaces. 

In the framework of the developed approach the dependencies of the critical Josephson current between superconducting leads of different geometries on temperature and applied magnetic field are investigated. It is shown that for experimentally relevant situation of rather short Josephson junction, which is studied in the companion paper \cite{Polevoy2025_joint}, the temperature dependence of the critical current varies from nearly linear with further saturation with decreasing temperature to faster than linear with further saturation. The particular temperature behavior depends on the model and on the geometry of the considered JJ. The dependence of the critical current on the magnetic field varies from the typical Fraunhofer behavior for wide JJs to the monotonic decrease for the narrow JJs, which is similar to the dependence typical for the diffusive, and not to the ballistic, sandwich JJs.

The Josephson transport through complex planar multi-terminal Josephson structures, in particular the dependencies of the critical Josephson current on temperature and the applied magnetic field,  is also studied. The results are in agreement with the experimental data \cite{Polevoy2025_joint}.

\begin{acknowledgments}
We thank Sir A. Geim for fruitful discussions. The work is supported by a grant RSF No 25-72-10127 \href{https://rscf.ru/project/25-72-10127/}{https://rscf.ru/project/25-72-10127/}.   
\end{acknowledgments}

\bibliography{SNSballistic}

@article{vanHeck2014,
  title = {Single fermion manipulation via superconducting phase differences in multiterminal Josephson junctions},
  author = {van Heck, B. and Mi, S. and Akhmerov, A. R.},
  journal = {Phys. Rev. B},
  volume = {90},
  issue = {15},
  pages = {155450},
  numpages = {9},
  year = {2014},
  month = {Oct},
  publisher = {American Physical Society},
  doi = {10.1103/PhysRevB.90.155450},
  url = {https://link.aps.org/doi/10.1103/PhysRevB.90.155450}
}

@Article{Riwar2016,
author={Riwar, Roman-Pascal
and Houzet, Manuel
and Meyer, Julia S.
and Nazarov, Yuli V.},
title={Multi-terminal Josephson junctions as topological matter},
journal={Nature Communications},
year={2016},
month={Apr},
day={04},
volume={7},
number={1},
pages={11167},
issn={2041-1723},
doi={10.1038/ncomms11167},
url={https://doi.org/10.1038/ncomms11167}
}

@article{Pfeffer2014,
  title = {Subgap structure in the conductance of a three-terminal Josephson junction},
  author = {Pfeffer, A. H. and Duvauchelle, J. E. and Courtois, H. and M\'elin, R. and Feinberg, D. and Lefloch, F.},
  journal = {Phys. Rev. B},
  volume = {90},
  issue = {7},
  pages = {075401},
  numpages = {8},
  year = {2014},
  month = {Aug},
  publisher = {American Physical Society},
  doi = {10.1103/PhysRevB.90.075401},
  url = {https://link.aps.org/doi/10.1103/PhysRevB.90.075401}
}

@Article{Strambini2016,
author={Strambini, E.
and D'Ambrosio, S.
and Vischi, F.
and Bergeret, F. S.
and Nazarov, Yu. V.
and Giazotto, F.},
title={The $\omega$-SQUIPT as a tool to phase-engineer Josephson topological materials},
journal={Nature Nanotechnology},
year={2016},
month={Dec},
day={01},
volume={11},
number={12},
pages={1055-1059},
issn={1748-3395},
doi={10.1038/nnano.2016.157},
url={https://doi.org/10.1038/nnano.2016.157}
}

@article{Pankratova2020,
  title = {Multiterminal Josephson Effect},
  author = {Pankratova, Natalia and Lee, Hanho and Kuzmin, Roman and Wickramasinghe, Kaushini and Mayer, William and Yuan, Joseph and Vavilov, Maxim G. and Shabani, Javad and Manucharyan, Vladimir E.},
  journal = {Phys. Rev. X},
  volume = {10},
  issue = {3},
  pages = {031051},
  numpages = {12},
  year = {2020},
  month = {Sep},
  publisher = {American Physical Society},
  doi = {10.1103/PhysRevX.10.031051},
  url = {https://link.aps.org/doi/10.1103/PhysRevX.10.031051}
}

@article{Gavensky2023,
doi = {10.1209/0295-5075/acb2f6},
url = {https://dx.doi.org/10.1209/0295-5075/acb2f6},
year = {2023},
month = {jan},
publisher = {EDP Sciences, IOP Publishing and Società Italiana di Fisica},
volume = {141},
number = {3},
pages = {36001},
author = {Peralta Gavensky, Lucila and Usaj, Gonzalo and Balseiro, C. A.},
title = {Multi-terminal Josephson junctions: A road to topological flux networks},
journal = {Europhysics Letters}
}

@article{Golubov2004,
  title = {The current-phase relation in Josephson junctions},
  author = {Golubov, A. A. and Kupriyanov, M. Yu. and Il'ichev, E.},
  journal = {Rev. Mod. Phys.},
  volume = {76},
  issue = {2},
  pages = {411--469},
  numpages = {0},
  year = {2004},
  month = {Apr},
  publisher = {American Physical Society},
  doi = {10.1103/RevModPhys.76.411},
  url = {https://link.aps.org/doi/10.1103/RevModPhys.76.411}
}

@article{Soloviev2021,
  title = {Miniaturization of Josephson Junctions for Digital Superconducting Circuits},
  author = {Soloviev, I.I. and Bakurskiy, S.V. and Ruzhickiy, V.I. and Klenov, N.V. and Kupriyanov, M.Yu. and Golubov, A.A. and Skryabina, O.V. and Stolyarov, V.S.},
  journal = {Phys. Rev. Appl.},
  volume = {16},
  issue = {4},
  pages = {044060},
  numpages = {12},
  year = {2021},
  month = {Oct},
  publisher = {American Physical Society},
  doi = {10.1103/PhysRevApplied.16.044060},
  url = {https://link.aps.org/doi/10.1103/PhysRevApplied.16.044060}
}

@Article{Ruzhickiy2023,
AUTHOR = {Ruzhickiy, Vsevolod and Bakurskiy, Sergey and Kupriyanov, Mikhail and Klenov, Nikolay and Soloviev, Igor and Stolyarov, Vasily and Golubov, Alexander},
TITLE = {Contribution of Processes in SN Electrodes to the Transport Properties of SN-N-NS Josephson Junctions},
JOURNAL = {Nanomaterials},
VOLUME = {13},
YEAR = {2023},
NUMBER = {12},
ARTICLE-NUMBER = {1873},
URL = {https://www.mdpi.com/2079-4991/13/12/1873},
PubMedID = {37368303},
ISSN = {2079-4991},
DOI = {10.3390/nano13121873}
}

@article{Bosboom2021,
title = "Selfconsistent 3D model of SN-N-NS Josephson junctions",
author = "V. Bosboom and {van der Vegt}, J.J.W. and {Yu Kupriyanov}, M. and A.A. Golubov",
note = "Publisher Copyright: {\textcopyright} 2021 The Author(s). Published by IOP Publishing Ltd.",
year = "2021",
month = nov,
day = "1",
doi = "10.1088/1361-6668/ac2d79",
volume = "34",
journal = "Superconductor science and technology",
issn = "0953-2048",
publisher = "Institute of Physics (IOP)",
number = "11",
}

@Article{Marychev2020,
author="Pavel M. Marychev and Denis Yu. Vodolazov",
title="A Josephson junction based on a highly disordered superconductor/low-resistivity normal metal bilayer",
journal="Beilstein Journal of Nanotechnology",
year="2020",
volume="11",
pages="858-865",
issn="2190-4286",
doi="10.3762/bjnano.11.71",
copyright="Marychev and Vodolazov; licensee Beilstein-Institut",
publisher="Beilstein-Institut",
URL="https://doi.org/10.3762/bjnano.11.71",
}

@article{2024_Bakurskiy,
author = {Sergey Bakurskiy and Vsevolod Ruzhickiy and Alexey Neilo and Nikolay Klenov and Igor Soloviev and Anna Elistratova and Andrey Shishkin and Vasily Stolyarov and Mikhail Kupriyanov},
title = {Thouless energy in Josephson SN-N-NS bridges},
journal = {Mesoscience \& Nanotechnology},
year = {2024},
volume = {1},
publisher = {Treatise LLC},
month = {Jul},
url = {https://jmsn.press/publications/10.64214/jmsn.01.01003},
number = {1},
doi = {10.64214/jmsn.01.01003}
}

@article{Polevoy_2026,
  title = {Proximity-effect engineering in aluminum-based planar Josephson junctions with intrinsic superconductivity},
  author = {Polevoy, K.B. and Bakurskiy, S.V. and Ruzhickiy, V.I. and Egorov, S.V. and Shishkin, A.G. and Frolov, A.S. and Kirsanova, M.A. and Krupatin, I.N. and Yanilkin, A.V. and Klenov, N.V. and Soloviev, I.I. and Golubov, A.A. and Kupriyanov, M.Yu. and Stolyarov, V.S.},
  journal = {Phys. Rev. Appl.},
  volume = {25},
  issue = {2},
  pages = {024030},
  numpages = {12},
  year = {2026},
  month = {Feb},
  publisher = {American Physical Society},
  doi = {10.1103/s1k7-wvw2},
  url = {https://link.aps.org/doi/10.1103/s1k7-wvw2}
}

@article{Vodolazov_2026,
  title = {Critical current oscillations and sign-reversal diode effect in planar SNS Josephson junctions: Manifestation of interface vortices},
  author = {Vodolazov, D. Yu. and Polevoy, K. B. and Shishkin, A. G. and Egorov, S. V. and Kuznetsova, P. R. and Elistratova, A. A. and Bakurskiy, S. V. and Stolyarov, V. S.},
  journal = {Phys. Rev. B},
  volume = {113},
  issue = {18},
  pages = {184509},
  numpages = {6},
  year = {2026},
  month = {May},
  publisher = {American Physical Society},
  doi = {10.1103/r628-qp67},
  url = {https://link.aps.org/doi/10.1103/r628-qp67}
}

@article{Bakurskiy_2026,
  title = {Direct evidence of the quasiparticle conversion length in proximity Josephson junctions},
  author = {Bakurskiy, S. V. and Skryabina, O. V. and Ruzhickiy, V. I. and Elistratova, A. A. and Polevoy, K. B. and Kuznetsova, P. R. and Shishkin, A. G. and Klenov, N. V. and Soloviev, I. I. and Kupriyanov, M. Yu. and Golubov, A. A. and Stolyarov, V. S.},
  journal = {Phys. Rev. B},
  volume = {113},
  issue = {13},
  pages = {134509},
  numpages = {10},
  year = {2026},
  month = {Apr},
  publisher = {American Physical Society},
  doi = {10.1103/9xbq-t64b},
  url = {https://link.aps.org/doi/10.1103/9xbq-t64b}
}

@article{Vischi2017,
  title = {Coherent transport properties of a three-terminal hybrid superconducting interferometer},
  author = {Vischi, F. and Carrega, M. and Strambini, E. and D'Ambrosio, S. and Bergeret, F. S. and Nazarov, Yu. V. and Giazotto, F.},
  journal = {Phys. Rev. B},
  volume = {95},
  issue = {5},
  pages = {054504},
  numpages = {10},
  year = {2017},
  month = {Feb},
  publisher = {American Physical Society},
  doi = {10.1103/PhysRevB.95.054504},
  url = {https://link.aps.org/doi/10.1103/PhysRevB.95.054504}
}

@article{Cohen2018,
author = {Yonatan Cohen  and Yuval Ronen  and Jung-Hyun Kang  and Moty Heiblum  and Denis Feinberg  and Régis Mélin  and Hadas Shtrikman },
title = {Nonlocal supercurrent of quartets in a three-terminal Josephson junction},
journal = {Proceedings of the National Academy of Sciences},
volume = {115},
number = {27},
pages = {6991-6994},
year = {2018},
doi = {10.1073/pnas.1800044115},
URL = {https://www.pnas.org/doi/abs/10.1073/pnas.1800044115},
eprint = {https://www.pnas.org/doi/pdf/10.1073/pnas.1800044115},
}

@article{Graziano2020,
  title = {Transport studies in a gate-tunable three-terminal Josephson junction},
  author = {Graziano, Gino V. and Lee, Joon Sue and Pendharkar, Mihir and Palmstr\o{}m, Chris J. and Pribiag, Vlad S.},
  journal = {Phys. Rev. B},
  volume = {101},
  issue = {5},
  pages = {054510},
  numpages = {7},
  year = {2020},
  month = {Feb},
  publisher = {American Physical Society},
  doi = {10.1103/PhysRevB.101.054510},
  url = {https://link.aps.org/doi/10.1103/PhysRevB.101.054510}
}

@Article{Graziano2022,
author={Graziano, Gino V.
and Gupta, Mohit
and Pendharkar, Mihir
and Dong, Jason T.
and Dempsey, Connor P.
and Palmstr{\o}m, Chris
and Pribiag, Vlad S.},
title={Selective control of conductance modes in multi-terminal Josephson junctions},
journal={Nature Communications},
year={2022},
month={Oct},
day={08},
volume={13},
number={1},
pages={5933},
issn={2041-1723},
doi={10.1038/s41467-022-33682-2},
url={https://doi.org/10.1038/s41467-022-33682-2}
}

@Article{Kozler2023,
AUTHOR = {Kolzer, Jonas and Jalil, Abdur Rehman and Rosenbach, Daniel and Arndt, Lisa and Mussler, Gregor and Schüffelgen, Peter and Grutzmacher, Detlev and Luth, Hans and Schapers, Thomas},
TITLE = {Supercurrent in Bi4Te3 Topological Material-Based Three-Terminal Junctions},
JOURNAL = {Nanomaterials},
VOLUME = {13},
YEAR = {2023},
NUMBER = {2},
ARTICLE-NUMBER = {293},
URL = {https://www.mdpi.com/2079-4991/13/2/293},
PubMedID = {36678045},
ISSN = {2079-4991},
DOI = {10.3390/nano13020293}
}

@Article{Draelos2019,
author={Draelos, Anne W.
and Wei, Ming-Tso
and Seredinski, Andrew
and Li, Hengming
and Mehta, Yash
and Watanabe, Kenji
and Taniguchi, Takashi
and Borzenets, Ivan V.
and Amet, Fran{\c{c}}ois
and Finkelstein, Gleb},
title={Supercurrent Flow in Multiterminal Graphene Josephson Junctions},
journal={Nano Letters},
year={2019},
month={Feb},
day={13},
publisher={American Chemical Society},
volume={19},
number={2},
pages={1039-1043},
issn={1530-6984},
doi={10.1021/acs.nanolett.8b04330},
url={https://doi.org/10.1021/acs.nanolett.8b04330}
}

@Article{Arnault2021,
author={Arnault, Ethan G.
and Larson, Trevyn F. Q.
and Seredinski, Andrew
and Zhao, Lingfei
and Idris, Sara
and McConnell, Aeron
and Watanabe, Kenji
and Taniguchi, Takashi
and Borzenets, Ivan
and Amet, Fran{\c{c}}ois
and Finkelstein, Gleb},
title={Multiterminal Inverse AC Josephson Effect},
journal={Nano Letters},
year={2021},
month={Nov},
day={24},
publisher={American Chemical Society},
volume={21},
number={22},
pages={9668-9674},
issn={1530-6984},
doi={10.1021/acs.nanolett.1c03474},
url={https://doi.org/10.1021/acs.nanolett.1c03474}
}

@Article{Huang2022,
author={Huang, Ko-Fan
and Ronen, Yuval
and M{\'e}lin, R{\'e}gis
and Feinberg, Denis
and Watanabe, Kenji
and Taniguchi, Takashi
and Kim, Philip},
title={Evidence for 4e charge of Cooper quartets in a biased multi-terminal graphene-based Josephson junction},
journal={Nature Communications},
year={2022},
month={May},
day={31},
volume={13},
number={1},
pages={3032},
issn={2041-1723},
doi={10.1038/s41467-022-30732-7},
url={https://doi.org/10.1038/s41467-022-30732-7}
}

@article{Kuprianov1986,
  title={Stationary properties of quasi-two-dimensional Josephson weak links},
  author={Kupriyanov, M Yu and Lukichev, V F and Orlikovskii, A A},
  journal={Microelectronics (in Russian)},
  volume={15},
  number={4},
  pages={328},
  year={1986},
  }

@article{Kuprianov1988,
  title={Influence of proximity effect in electrodes and interface transparency on stationary properties of clean Josephson SNS structures},
  author={Kupriyanov, M Yu and Lukichev, V F},
  journal={Proceedings of the Institute of General Physics of RAS (in Russian)},
  volume={14},
  pages={160},
  year={1988},
  }

@article{Eilenberger1968,
author={Eilenberger, Gert},
title={Transformation of Gorkov's equation for type II superconductors into transport-like equations},
journal={Zeitschrift f{\"u}r Physik A Hadrons and nuclei},
year={1968},
month={Apr},
day={01},
volume={214},
number={2},
pages={195-213},
issn={0939-7922},
doi={10.1007/BF01379803},
}

@article{Larkin1968,
  title={Quasi-classical method in the theory of superconductivity},
  author={Larkin, AI and Ovchinnikov, Yu N},
  journal={Zh. Eksp. Teor. Fiz.},
  volume={55},
  number={6},
  year={1968},
  note = {[Sov. Phys. JETP \textbf{28}, 1200 (1969)  },
}

@article{Zaitsev1984,
  title={Quasiclassical equations of the theory of superconductivity for contiguous metals and the properties of constricted microcontacts},
  author={Zaitsev, A V},
  journal={Sov. Phys. JETP},
  volume={59},
  number={1015},
  year={1984},
  }

@article{Barzykin1999,
title = {Coherent transport and nonlocality in mesoscopic SNS junctions: anomalous magnetic interference patterns},
journal = {Superlattices and Microstructures},
volume = {25},
number = {5},
pages = {797-807},
year = {1999},
issn = {0749-6036},
doi = {https://doi.org/10.1006/spmi.1999.0731},
url = {https://www.sciencedirect.com/science/article/pii/S0749603699907310},
author = {Victor Barzykin and Alexandre M. Zagoskin},
}

@article{Yokoyama2015,
  title = {Singularities in the Andreev spectrum of a multiterminal Josephson junction},
  author = {Yokoyama, Tomohiro and Nazarov, Yuli V.},
  journal = {Phys. Rev. B},
  volume = {92},
  issue = {15},
  pages = {155437},
  numpages = {10},
  year = {2015},
  month = {Oct},
  publisher = {American Physical Society},
  doi = {10.1103/PhysRevB.92.155437},
  url = {https://link.aps.org/doi/10.1103/PhysRevB.92.155437}
}

@article{Polevoy2025_joint,
  title = {Multiterminal Quasiballistic Josephson Effect in Monocrystalline Gold},
  author = {Polevoy, K. B. and Bobkov, G. A. and Kalashnikov, D. S. and Shishkin, A. G. and Trofimov, I. V. and Bobkov, A. M. and Sedov, E. A. and Krivko, E. A. and Tarkhov, M. A. and Bobkova, I. V. and Stolyarov, V. S.},
  journal = {Phys. Rev. Lett.},
  volume = {137},
  issue = {16},
  pages = {166001},
  numpages = {7},
  year = {2026},
  month = {Oct},
  publisher = {American Physical Society},
  doi = {10.1103/rp26-lscg},
  url = {https://link.aps.org/doi/10.1103/rp26-lscg}
}

@ARTICLE{Kulik1969,
       author = {{Kulik}, I.~O.},
        title = "{Macroscopic Quantization and the Proximity Effect in S-N-S Junctions}",
      journal = {Soviet Journal of Experimental and Theoretical Physics},
         year = 1970,
        month = jan,
       volume = {30},
        pages = {944},
       adsurl = {https://ui.adsabs.harvard.edu/abs/1969JETP...30..944K}
}

@article{Ishii1970,
    author = {Ishii, Chikara},
    title = {Josephson Currents through Junctions with Normal Metal Barriers},
    journal = {Progress of Theoretical Physics},
    volume = {44},
    number = {6},
    pages = {1525-1547},
    year = {1970},
    month = {12},
    issn = {0033-068X},
    doi = {10.1143/PTP.44.1525},
    url = {https://doi.org/10.1143/PTP.44.1525},
    eprint = {https://academic.oup.com/ptp/article-pdf/44/6/1525/5378760/44-6-1525.pdf},
}

@article{Bardeen1972,
  title = {Josephson Current Flow in Pure Superconducting-Normal-Superconducting Junctions},
  author = {Bardeen, John and Johnson, Jared L.},
  journal = {Phys. Rev. B},
  volume = {5},
  issue = {1},
  pages = {72--78},
  numpages = {0},
  year = {1972},
  month = {Jan},
  publisher = {American Physical Society},
  doi = {10.1103/PhysRevB.5.72},
  url = {https://link.aps.org/doi/10.1103/PhysRevB.5.72}
}

@article{Kupriyanov1981,
    author = {Kupriyanov, M. Yu.},
    title = {Stationary properties of clean SNS sandwiches},
    journal = {Soviet Journal Low Temperature Physics},
    volume = {7},
    number = {6},
    pages = {342-345},
    year = {1981},
    month = {06},
    issn = {0360-0335},
    doi = {10.1063/10.0030532},
    url = {https://doi.org/10.1063/10.0030532},
    eprint = {https://pubs.aip.org/aip/ltp/article-pdf/7/6/342/20299296/342\_1\_10.0030532.pdf},
}

@article{Galaktionov2002,
  title = {Quantum interference and supercurrent in multiple-barrier proximity structures},
  author = {Galaktionov, Artem V. and Zaikin, Andrei D.},
  journal = {Phys. Rev. B},
  volume = {65},
  issue = {18},
  pages = {184507},
  numpages = {13},
  year = {2002},
  month = {Apr},
  publisher = {American Physical Society},
  doi = {10.1103/PhysRevB.65.184507},
  url = {https://link.aps.org/doi/10.1103/PhysRevB.65.184507}
}

@article{Eschrig2009,
  title = {Scattering problem in nonequilibrium quasiclassical theory of metals and superconductors: General boundary conditions and applications},
  author = {Eschrig, Matthias},
  journal = {Phys. Rev. B},
  volume = {80},
  issue = {13},
  pages = {134511},
  numpages = {22},
  year = {2009},
  month = {Oct},
  publisher = {American Physical Society},
  doi = {10.1103/PhysRevB.80.134511},
  url = {https://link.aps.org/doi/10.1103/PhysRevB.80.134511}
}

@article{Haberkorn1978,
author = {Haberkorn, W. and Knauer, H. and Richter, J.},
title = {A theoretical study of the current-phase relation in Josephson contacts},
journal = {physica status solidi (a)},
volume = {47},
number = {2},
pages = {K161-K164},
doi = {https://doi.org/10.1002/pssa.2210470266},
url = {https://onlinelibrary.wiley.com/doi/abs/10.1002/pssa.2210470266},
eprint = {https://onlinelibrary.wiley.com/doi/pdf/10.1002/pssa.2210470266},
year = {1978}
}

@article{Cuevas2007,
  title = {Magnetic Interference Patterns and Vortices in Diffusive SNS Junctions},
  author = {Cuevas, J. C. and Bergeret, F. S.},
  journal = {Phys. Rev. Lett.},
  volume = {99},
  issue = {21},
  pages = {217002},
  numpages = {4},
  year = {2007},
  month = {Nov},
  publisher = {American Physical Society},
  doi = {10.1103/PhysRevLett.99.217002},
  url = {https://link.aps.org/doi/10.1103/PhysRevLett.99.217002}
}

@article{Ledermann1999,
  title = {Nonlocality in mesoscopic Josephson junctions with strip geometry},
  author = {Ledermann, Urs and Fauch\`ere, Alban L. and Blatter, Gianni},
  journal = {Phys. Rev. B},
  volume = {59},
  issue = {14},
  pages = {R9027--R9030},
  numpages = {0},
  year = {1999},
  month = {Apr},
  publisher = {American Physical Society},
  doi = {10.1103/PhysRevB.59.R9027},
  url = {https://link.aps.org/doi/10.1103/PhysRevB.59.R9027}
}

\end{document}